\documentclass{elsart}

\begin{document}


\runauthor{Aegerter}
\begin{frontmatter}

\title{Dynamic roughening of the magnetic flux landscape in YBa$_2$Cu$_3$O$_{7-x}$.}

\author{C.~M.~Aegerter\thanksref{label2}}
\author{M. S. Welling}
\author{R. J. Wijngaarden}
\address{Department of Physics and Astronomy; Vrije Universiteit; 1081HV
Amsterdam; The Netherlands.}
\thanks[label2]{Present address: Fachbereich Physik, Universit\"at Konstanz,
P.O. Box 5560, 78457 Konstanz, Germany.}

\begin{abstract}
We study the magnetic flux landscape in YBa$_2$Cu$_3$O$_{7-x}$
thin films as a two dimensional rough surface. The vortex density
in the superconductor forms a self-affine structure in both space
and time. This is characterized by a roughness exponent $\alpha =
0.76(3)$ and a growth exponent $\beta = 0.57(6)$. This is due to
the structure and distribution of flux avalanches in the
self-organized critical state, which is formed in the
superconductor. We also discuss our results in the context of
other roughening systems in the presence of quenched disorder.
\end{abstract}
\end{frontmatter}


\section{Introduction}

When a type II superconductor is put in a slowly ramped external
field, magnetic vortices start to penetrate the sample from its
edges. These vortices get pinned by dislocations or other
crystallographic defects, leading to the build-up of a flux
gradient, which is only marginally stable, as is the slope of a
slowly grown pile of sand \cite{degen}. Thus it can happen that
small changes in the applied field can lead to large
rearrangements of flux in the sample, known as flux avalanches
\cite{ernesto}. Due to the analogy of the flux landscape with a
pile of sand, the properties of flux avalanches have long been
studied in the context of self-organized criticality (SOC)
\cite{SOC}, which predicts that many slowly driven non-equilibrium
systems have avalanches which are distributed according to a
power-law \cite{tang}. As a matter of fact, vortex avalanches in
superconductors are thought of as an ideal experimental system in
which to study SOC, due to the over-damped dynamics of the
vortices \cite{ernesto,frette}. In the past, power-law distributed
avalanches have been observed in many controlled experiments,
which were ascribed to SOC \cite{field}. Furthermore, the
microscopic dynamics of the particles (vortices) is well known
\cite{blatterbible} and the collective dynamics can then be for
instance studied in detail using molecular dynamics simulations
\cite{olson}. In more detail however, SOC predicts that a system
not only shows power-law behaviour, but that it organizes into a
critical state, and should thus show finite-size scaling in the
distribution of avalanches as well \cite{jensen}. This has now
also recently been shown for the flux-avalanches in a thin film of
YBa$_2$Cu$_3$O$_{7-x}$ (YBCO) \cite{socsup}, where also the shape
of the flux avalanches and their fractal dimension was
characterised. The shape and distribution of avalanches moreover
strongly influence the shape of the magnetic flux landscape,
leading to a rough, self-affine surface \cite{chen}. The
characteristic exponents of this surface can then be obtained
quantitatively from the avalanche properties via a set of scaling
relations derived by Paczuski {\em et al.} for many SOC models
\cite{pacz}.

Here, we study the roughening properties of the magnetic flux
landscape in a thin film of YBCO in two dimensions (2d). While
dynamic roughening has been experimentally studied in many 1d
systems \cite{bact,porous,paper}, among which there was also a
study of the front of penetrating flux in YBCO \cite{radu}, 2d
characterization of roughening systems are rare in the
experimental literature \cite{ricepile}. A full 2d
characterisation of the roughness properties makes it possible to
compare properties of the avalanches with those of the surface, in
order to have a stringent test for SOC in the flux avalanches in
YBCO \cite{socsup}. Furthermore, we compare the roughness results
with numerical integrations of the Edwards-Wilkinson (EW) equation
\cite{ew} in the presence of quenched disorder in order to have a
comparison with the static pinning in the experiment.

Section 2 will describe the experimental setup in detail, while
section 3 will introduce the analysis methods for a 2d surface,
which are applied to the data. In section 4, we present the
results of the numerical studies of the quenched EW equation,
before turning to the roughness of the flux landscape in section
5. These roughness results will finally be compared with the
avalanche properties, determined elsewhere \cite{socsup}, in
section 6.

\section{Experimental Setup}

The magnetic flux density $B_z$ just above the YBCO thin film is
measured by means of the Faraday-effect \cite{mo} in an advanced
magneto-optical microscope \cite{arvoo}. In this setup, the
polarization rotation angle is measured directly by means of a
lock-in technique. The YBCO films studied here were grown on a
NdGaO$_3$ substrate to a thickness of 80 nm using pulsed laser
ablation \cite{sample}. Pinning sites in the sample consist mostly
of screw dislocations and are distributed uniformly over the
sample, acting as point pins \cite{defects}.

\begin{figure}[hbt]
\input{epsf}
\epsfxsize 10cm
\centerline{\epsfbox{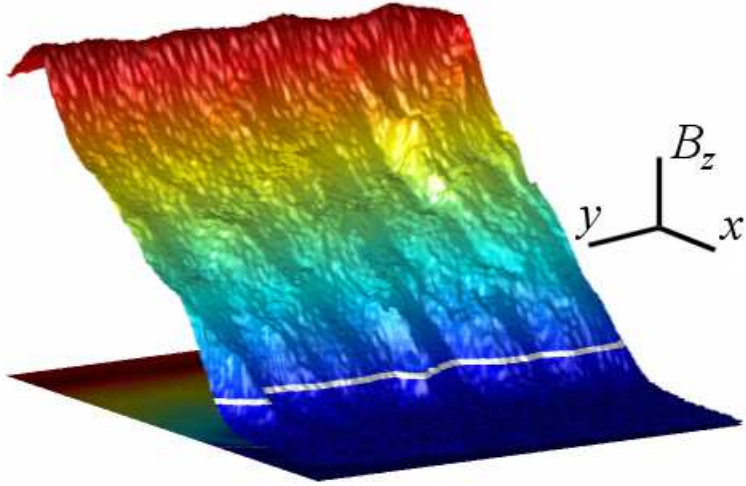}} \caption[~]{The magnetic flux
surface at an applied field of 12 mT in the YBCO thin film. As can
be seen, when the flux density $B_z$ is plotted as height, a shape
reminiscent of a pile is obtained, which has a rough surface. The
self-affine properties of the fluctuations around the mean surface
are analysed below, after subtracting the average profile.}
\label{mo}
\end{figure}

The sample is cooled in zero applied field to 4.2 K, after which
the field is slowly increased in steps of 50 $\mu$T, where at each
field the sample is allowed to relax for 10 s before an image is
taken with a high resolution charge coupled device camera (782
$\times$ 582 pixels) measuring the flux density $B_z (x,y)$. In
each experiment, 300 such field steps are taken leading to a
magnetic flux landscape as shown in Fig.~\ref{mo}. In the
analysis, only the last 140 images of each run were used in order
to have a flux landscape spanning more than 100 pixels. The
results below come from a total of five such experiments on the
same sample.

\section{Analysis Methods}
\label{anal}

As can be seen from Fig.~\ref{mo}, the magnetic flux surface,
$B_z(x,y)$, has an average profile, $\langle B_z(y)\rangle_x$,
with high values near the sample edge and zero magnetic flux
inside the sample, where $\langle ^. \rangle_s$ denotes an average
over the subscript $s$. In order to only study the properties of
the fluctuations in the surface, this average profile is
subtracted from the data, such that in the following the
properties of $b(x,y) = B_z(x,y) - \langle B_z(y)\rangle_x$ are
studied. Furthermore, the time evolution of the flux landscape is
investigated by choosing the area which is studied to start at the
mean position of the front of penetrating flux (indicated by the
white line in Fig.~\ref{mo}) and going backwards towards the
sample edge for 128 pixels. In this way a stationary state is
mimicked where the height of the flux landscape is approximately
constant in time. The level of the flux front is chosen to be
about three times the standard deviation of the noise in the
Meissner phase, such that an accurate determination is possible.

The roughness of an interface is classically quantified via a
determination of the width of the interface, as given by the
second moment, $w(t,L)$ of the fluctuations around the mean
interface \cite{barabasi}. For a roughening system characterized
by a self-affine structure in space and time, the width increases
with time as a power-law, $w(t) \propto t^\beta$, with the growth
exponent $\beta$. At long times, after the correlation length has
reached the system size, $L$, the width saturates at a value
$w_{sat}(L) \propto L^\alpha$, depending on the system size again
as a power-law \cite{barabasi}. Here, $\alpha$ is called the
roughness exponent. In experiments, it is often more useful to
determine the characteristic exponents from the correlation
functions rather than from the interface width \cite{power}. The
correlation function, in 2d, is given by:
\begin{equation}
C^2(x,y,t) = \langle (h(\xi + x,\eta + y,\tau + t) -
h(\xi,\eta,\tau))^2\rangle_{\xi,\eta,\tau}.
\end{equation}
In contrast to the width, the correlation function is obtained
averaging over more points, leading to a more reliable estimate.
The characteristic exponents can still be determined from the
correlation function, since for a self-affine structure the
scaling behaviour is as that of the width, i.e. $C(r,0) \propto
r^\alpha$ and $C(0,t) \propto t^\beta$ \cite{barabasi}. Here, $r =
\sqrt{x^2 + y^2}$ is the radial distance in the plane.

Experimental noise in the apparatus (e.g. photon counting noise)
may hamper the determination of the correlation function, such
that power-law behaviour is lost at small scales. In our
experiment, the flux surface $b(x,y)$ is in fact determined by
contributions from both the ideal flux surface, $h(x,y)$ and
experimental noise, predominantly due to photon counting noise,
$\epsilon(x,y)$. Inserting a height fluctuation with an added
noise component, $b(x,y) = h(x,y) + \epsilon(x,y)$ into the
definition of the correlation function, one finds that the
behaviour for the ideal fluctuations can still be obtained from
the data, as long as the properties of the noise are known:
\begin{equation}
C^2_{h}(x,y,t) = C^2_{b}(x,y,t) - 2\sigma^2(x,y). \label{noise}
\end{equation}
Here $\sigma(x,y)$ is the strength of the noise as given by
$\langle \epsilon^2(x,y) \rangle$. In our system, we can
experimentally determine the properties of the noise, such that a
correction for the noise according to Eq.~\ref{noise} and hence
proper determination of the exponents is possible \cite{noise}.

\section{Numerical Simulations}

In order to study the influence of quenched disorder, analogous to
the static point pins in the superconductor, on the roughening
properties of the flux surface, we also carried out numerical
simulations of the two dimensional EW equation in the presence of
static disorder. The EW equation describes the height of an
interface, $h(x,y,t)$, as a function of time, which is pulled
though a disordered medium with a speed $v$. In the experimental
case, the driving can be seen as the increasing applied magnetic
\begin{figure}[hbt]
\input{epsf}
\epsfxsize 10cm
\centerline{\epsfbox{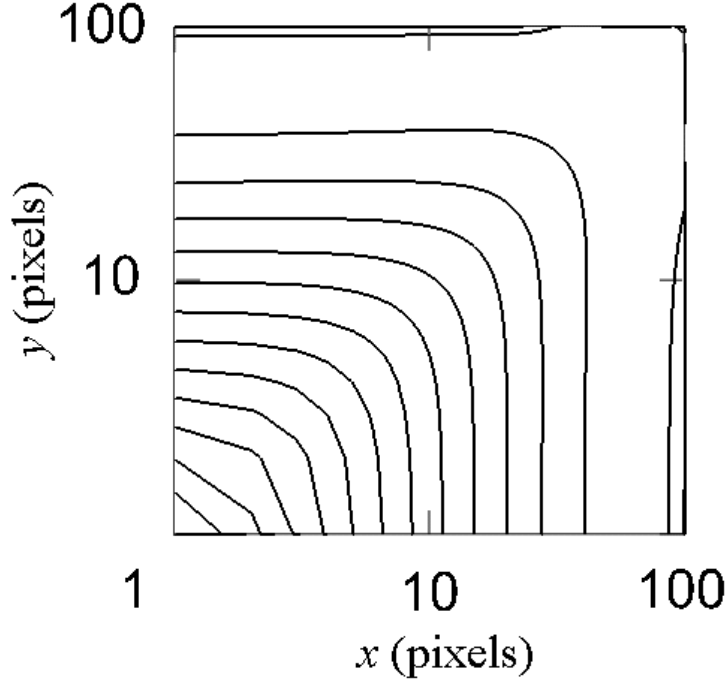}} \caption[~]{Double logarithmic
plot of the full 2d correlation function of the height
fluctuations in the numerical simulation of the quenched EW
equation. The contours are logarithmically spaced, such that
equidistant contours indicate power-law behaviour. Furthermore, on
a linear scale the contours are circular, indicating the isotropic
nature of the system in the $x$- and $y$-directions (see also the
inset of Fig.~\ref{ewcr}).} \label{ewcx}
\end{figure}
field $H_{ext}$. In this case, the roughening of the interface is
due to the disorder, $\eta(x,y,t)$, which is usually considered to
have a white spectrum in both space and time. Furthermore, there
is an elastic term, which leads to the smoothing of the surface
\cite{ew}.
\begin{equation}
\frac{\partial h (x,y,t)}{\partial t} = D
\left(\left(\frac{\partial}{\partial x}\right)^2 +
\left(\frac{\partial}{\partial y}\right)^2\right) h (x,y,t) +
\eta(x,y,t) + v. \label{EW}
\end{equation}
In the case of a white spectrum of the disorder, the equation can
be solved exactly and in one dimension a self-affine interface is
obtained, whereas in two dimensions roughening is only marginal
\cite{barabasi}. However, in the presence of a combination of
quenched and dynamic disorder, the equation can no longer be
solved analytically and numerical simulations show roughness also
in 2d \cite{halpin}. Here, we have numerically integrated the EW
equation on a lattice of 128$\times$128 pixels with periodic
boundary conditions. In order to have both quenched and thermal
disorder in the model, we add two disorder terms to the elastic
term in Eq.~\ref{EW}, $\eta_q(x,y)$ and $\eta_d(x,y,t)$, where
$\eta_q$ is constant in time and represents the quenched part of
the disorder and $\eta_d$ has a white spectrum in time and
represents the dynamic disorder. In the presence of quenched
disorder we speak of the quenched EW equation. Both disorder terms
have a white spectrum in space. The strength of both terms is
equal for the results presented below, however the values of the
exponents are very robust and ratios of 0.5 $< \eta_q / \eta_d <$
10 yield the same exponents within errors. The integration is run
for 10000 time steps until the width of the interface saturates
and the full 2d correlation function is determined subsequently in
20 different simulations. In Fig.~\ref{ewcx}, the spatial
component, $C(x,y)$, is shown in a logarithmic contour plot. Since
the contours are logarithmically spaced, evenly spaced contours in
the plot imply power-law behaviour. The fact that the contours are
circularly
\begin{figure}[hbt]
\input{epsf}
\epsfxsize 10cm
\centerline{\epsfbox{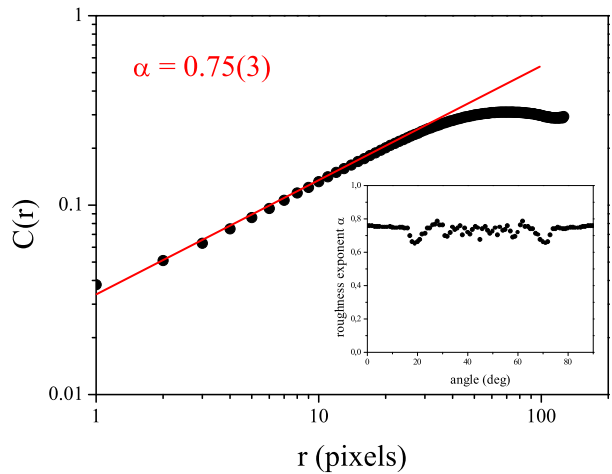}} \caption[~]{Radial average of the
full 2d correlation function of the height variations of the
numerical simulation of the quenched EW equation. The straight
line on the double logarithmic plot indicates power-law behaviour,
where the slope of the curve determines the roughness exponent of
$\alpha$ = 0.75(3). The inset shows the angular dependence of the
roughness exponent $\alpha$, as determined from a fit to different
radial projections of the full 2d correlation function in the
range of $r <$ 11 pixels. The values of $\alpha$ are independent
of the radial angle, as it should be for an isotropic system.}
\label{ewcr}
\end{figure}
shaped on a linear scale, indicates that the system is isotropic
as it should be. This can also be seen by directly fitting a
power-law dependence to radial projections of $C(x,y)$ over the
range of $r <$ 11 pixels. As can be seen in the inset to
Fig.~\ref{ewcr}, the roughness exponent, $\alpha$, does not depend
on the radial angle, such that a radial average, shown in the main
part of the figure can be used to obtain a good determination of
$\alpha$. As can be seen, the correlation function is linear in
the log-log plot over more than a decade with an exponent of
$\alpha$ = 0.75(3), indicated by the straight line.

\begin{figure}[hbt]
\input{epsf}
\epsfxsize 10cm
\centerline{\epsfbox{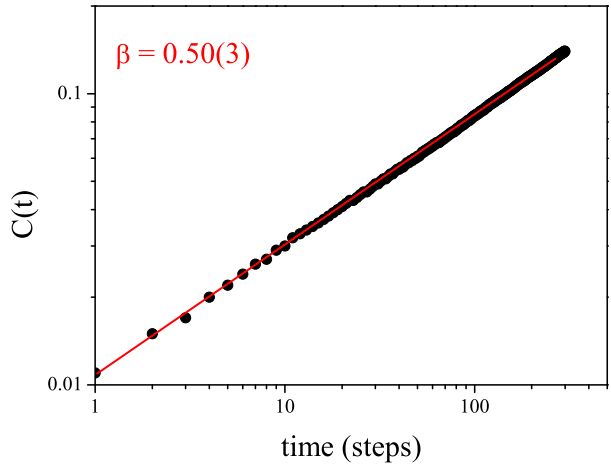}} \caption[~]{The time correlation
function for the numerical simulation of the quenched EW equation
on a double logarithmic plot. The straight line indicates that the
correlation function is a power-law with an exponent of $\beta$ =
0.50(3). } \label{ewct}
\end{figure}

The temporal behaviour of the quenched EW equation can be
characterized by the growth exponent $\beta$. This is determined
from the temporal correlation function $C(t)$, which is shown in
Fig.~\ref{ewct} on a double logarithmic scale. There is good
power-law behaviour in the time correlation function, with a
growth exponent of $\beta$ = 0.50(3), as determined from a linear
fit over 100 time steps in the double logarithmic plot.

Thus in the presence of quenched disorder, the roughness
properties of the EW equation change drastically with a strong
increase of both the roughness and the growth exponent (in the
absence of static disorder we obtain $\alpha \simeq \beta \simeq$
0.1). In the magnetic flux scape discussed below, the dynamics is
mainly governed by the static pinning landscape, such that a model
including quenched disorder is necessary in order describe the
main features of the system. Note also that one of the standard
sandpile models \cite{oslo}, which is used to describe the
dynamics of a rice-pile, has been mapped exactly onto the quenched
EW equation \cite{pruessner}.

\section{Flux landscape}

Given the rough landscape shown in Fig.~\ref{mo}, we determine the
self-affine properties of the fluctuations around the mean surface
as discussed in Sec.~\ref{anal} above.
\begin{figure}[hbt]
\input{epsf}
\epsfxsize 9cm
\centerline{\epsfbox{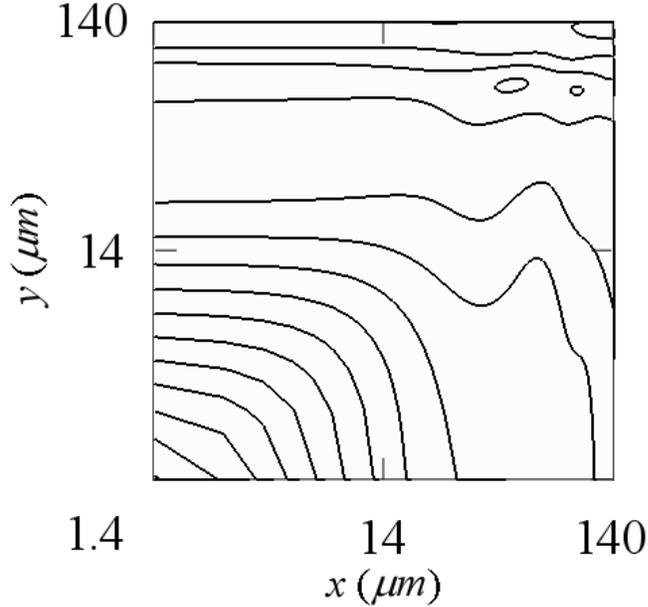}} \caption[~]{Contour plot of the
full 2d correlation function of the magnetic flux landscape. The
contours are logarithmically spaced, such that equidistant
contours imply a power-law dependence. The contours have a
circular shape on a linear scale, which indicates that the system
behaves isotropically in the $x$- and $y$-directions (see also
inset of Fig.~\ref{ybcocr}).} \label{ybcocx}
\end{figure}
In order to have a reasonable size for the 2d area used in the
analysis, we analyse images starting from an applied field of 5 mT
and determine the spatial and temporal correlation functions for
the subsequent 128 images. Averaged over these 128 images and over
all experiments, the full 2d correlation function, $C(x,y)$, is
shown as a logarithmic contour plot in Fig.~\ref{ybcocx}. As was
the case in Fig.~\ref{ewcx} above, the equidistant contours imply
power-law behaviour and circular contours on a linear scale
support the fact that the system is isotropic, such that a radial
average may be performed in order to determine the roughness
exponent more precisely. Another indication of of the isotropic
nature of the system can be seen in the inset of
\begin{figure}[hbt]
\input{epsf}
\epsfxsize 10cm
\centerline{\epsfbox{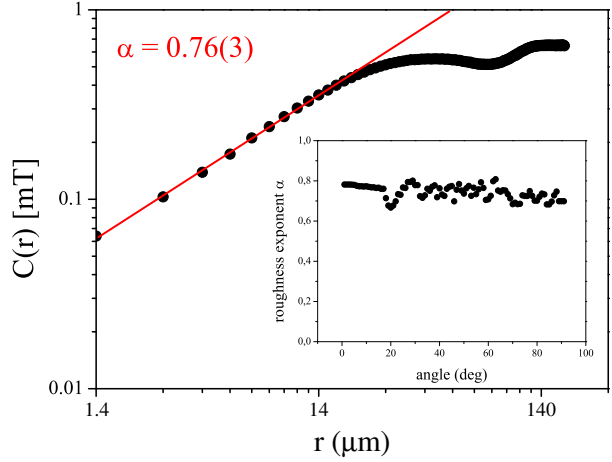}} \caption[~]{Radial average of
the correlation function of the magnetic flux surface. The inset
shows the angular dependence of the roughness exponent $\alpha$,
as obtained from a power-law fit to the full 2d correlation
function in the region $r <$ 35 $\mu m$. The roughness exponent is
independent of the radial angle, implying that the system is
isotropic and that a radial average may be performed in order to
determine $\alpha$.} \label{ybcocr}
\end{figure}
Fig.~\ref{ybcocr}, where the radial dependence of the roughness
exponent is shown, as determined from a power-law fit on a scale
$r <$ 35 $\mu m$. As the figure shows, the roughness exponent does
not depend on the radial angle, which means that a radial average
can safely be carried out. The corresponding radial average,
$C(r)$, is shown in the main part of Fig.~\ref{ybcocr}, after a
noise correction as described above \cite{noise}, on a double
logarithmic plot. As can be seen there is good power-law behaviour
over more than one decade with a roughness exponent of $\alpha$ =
0.76(3), indicated by the straight line in the figure. Again, the
value of the exponent is obtained from a linear fit to the data in
the double logarithmic plot in the range of $r <$ 35 $\mu m$. The
roughness exponent, which is obtained is in good agreement with
that from the numerical integration of the quenched EW equation
above.

\begin{figure}[hbt]
\input{epsf}
\epsfxsize 10cm
\centerline{\epsfbox{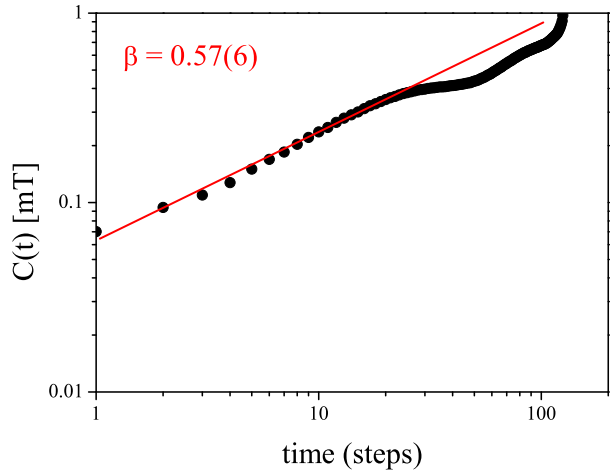}} \caption[~]{The time correlation
function for the magnetic flux surface. The straight line
indicates power-law behaviour, with an exponent of $\beta$ =
0.57(6). } \label{ybcoct}
\end{figure}

The temporal behaviour is characterized by the time correlation
function of the height fluctuations. Again, noise in the
experimental system distorts the data via an intrinsic width
\cite{noise}. After correction for the experimentally determined
noise, the data are plotted in Fig.~\ref{ybcoct}. As can be seen,
the correlation function is a power-law over a decade in time,
where the straight line in the figure indicates the exponent of
$\beta$ = 0.57(6). Here, the exponent was obtained from a linear
fit to the double logarithmic plot over the first 15 time steps.
Again, this is in agreement with the result from the integration
of the quenched EW equation. However, as mentioned above, the
surface roughness is closely connected with the avalanche
behaviour described by SOC \cite{pacz}. The avalanche properties
in our YBCO sample have been studied before, which allows a
detailed quantitative comparison of the roughness properties with
those of the avalanches \cite{socsup}. This is done in the next
section.

\section{Comparison with Avalanche Properties}
\label{aval} It has been noted that the roughness of a surface of
a SOC system is created by the shape and distribution of the
avalanches in the system \cite{chen}. Thus there should be a
connection between these two phenomena, which can be tested
\cite{pacz}. Such a quantitative connection has been previously
shown in the properties of the avalanches and the surface
roughness of a three dimensional pile of rice \cite{ricepile}.
From a theoretical point of view, the scaling relations have been
derived in the context of models of extremal dynamics as well as
the specific problem of a rice pile surface \cite{pacz,ricepile}.
When the avalanches have a fractal dimension of $D$, which can be
obtained from a finite-size scaling analysis of the avalanche size
distribution and the active area of the avalanches has a fractal
dimension of $d_B$, then the roughness exponent is given by
$\alpha = D - d_B$. Using the values previously determined in the
same YBCO samples \cite{socsup} for the avalanche properties, i.e.
$D$ = 1.89(3) and $d_B$ = 1.18(5), we obtain $\alpha$ = 0.71(6).
This is in good agreement with the value of $\alpha$ = 0.76(3)
determined above, as well as with the result from our numerical
integration of the quenched EW equation. The growth exponent can
be obtained via the dynamical exponent, $z = \alpha / \beta$,
which describes the scale dependence of the cross-over time
$t_\times$, at which the width of the surface saturates. This is
given by $z = D (2 - \tau)$, where $\tau$ is the exponent of the
avalanche size distribution. Again using the values from
Ref.~\cite{socsup}, $\tau$ = 1.29(2) and $D$ = 1.89(3), we obtain
$z$ = 1.34(4) and hence $\beta$ = 0.53(9). This is in good
agreement with the determination from the roughness analysis
above, $\beta$ = 0.57(6), as well as with the results from the
simulations of the quenched EW equation. Thus, not only can the
Oslo-model \cite{oslo} be mapped onto the quenched EW equation
\cite{pruessner}, also the more general scaling relations for SOC
models derived by Paczuski {\em et al.} \cite{pacz} can be checked
in a system described by the quenched EW equation. This gives
further support to the connection between SOC and roughening
physics \cite{alava,zapperi}, which has already been shown for the
surface of a pile of rice \cite{ricepile}.

\section{Conclusions}

In conclusion, we have shown that the two dimensional flux surface
of an YBCO thin film in the mixed state is self-affine and shows
power-law scaling in its roughness and growth, given by $\alpha =
0.76(3)$ and $\beta = 0.57(6)$. Furthermore, this behaviour is in
good agreement with the expectations from a simple roughening
system (the EW equation) in the presence of quenched disorder. In
addition, the surface roughness is connected to the avalanche
properties of the magnetic flux jumps, as is expected for a SOC
system. The scaling relations of Paczuski {\em et al.}
\cite{pacz}, can be used to quantitatively predict the values of
the roughness and growth exponents from the avalanche size
distribution exponent and dimensions. Using an earlier
characterisation of the avalanche properties in our sample
\cite{socsup}, we obtain good agreement between the expectation
from the scaling relations and the experimentally observed
roughness exponents. This shows that the flux landscape in YBCO is
formed by the avalanches of a SOC process. Also it shows the
intimate connection between the roughening of an interface and
avalanche dynamics, or SOC in general \cite{alava}.

\nonumber \section{Acknowledgements}

We would like to thank Jan Rector for providing the sample. This
work was supported by FOM (Stichting voor Fundamenteel Onderzoek
der Materie), which is financially supported by NWO (Nederlandse
Organisatie voor Wetenschappelijk Onderzoek).

\bibliographystyle{prsty}

\end{document}